\documentclass[prx, twocolumn,superscriptaddress,  notitlepage, showpacs, floatfix]{revtex4-1}
\usepackage{amsmath,bm}
\usepackage{graphicx}
\usepackage{amsfonts}
\usepackage{amssymb}
\usepackage{epstopdf}
\usepackage{bbold}
\usepackage[hypertex]{hyperref}
\usepackage{hyperref}
\usepackage{color}

\newcommand{\mat}[2]{
\left[\begin{array}{#1}
#2
\end{array}
\right]}

\DeclareMathAlphabet\mathbfcal{OMS}{cmsy}{b}{n}

\newcommand{\ket}[1]{| #1 \rangle}
\newcommand{\bra}[1]{\langle #1 |}

\newcommand{\proj}[1]{|#1 \rangle\!\langle #1 |}
\newcommand{\coh}[2]{|#1 \rangle\!\langle #2 |}
\newcommand{\ii}{\mathrm{i}}
\newcommand{\cE}{\mathcal{E}}
\newcommand{\cH}{\mathcal{H}}
\newcommand{\tr}{\mathrm{Tr}}
\renewcommand{\t}[1]{\mathrm{#1}}
\newcommand{\be}{\begin{equation}}
\newcommand{\ee}{\end{equation}}
\newcommand{\bi}{\chi}
\newcommand{\combs}{\Upsilon}

\begin{document}
\title{Adaptive quantum metrology under general Markovian noise}
\author{Rafa\l{} Demkowicz-Dobrza\'nski}
\affiliation{Faculty of Physics, University of Warsaw,  ul. Pasteura 5, PL-02-093 Warszawa, Poland}
\author{Jan Czajkowski}
\affiliation{Faculty of Physics, University of Warsaw,  ul. Pasteura 5, PL-02-093 Warszawa, Poland}
\affiliation{QuSoft, University of Amsterdam, Institute for Logic, Language and Computaion (ILLC),\\
P.O. Box 94242, 1090 GE Amsterdam, The Netherlands}
\author{Pavel Sekatski}
\affiliation{Institut fur Theoretische Physik, Universit\"{a}t Innsbruck, Technikerstr. 21a, A-6020 Innsbruck, Austria}

\begin{abstract}
We consider a general model of unitary parameter estimation in presence of Markovian noise,
where the parameter to be estimated is associated with the Hamiltonian part of the dynamics.
In absence of noise, unitary parameter can be estimated with precision scaling as $1/T$, where $T$ is the total probing time. We provide a simple algebraic condition involving solely the operators appearing in the quantum Master equation, implying at most $1/\sqrt{T}$ scaling of precision under the most general adaptive quantum estimation strategies. We also discuss the requirements a quantum error-correction like protocol must satisfy in order to regain the $1/T$ precision scaling in case the above mentioned algebraic condition is not satisfied.
Furthermore, we apply the developed methods to understand fundamental precision limits in atomic interferometry with many-body effects taken into account, shedding  new light on the performance of non-linear metrological models.
\end{abstract}
\maketitle

\section{Introduction}
With rapid advancements in quantum optical experimental techniques, the field of quantum metrology \cite{Giovannetti2011, Toth2014,  demkowicz2015quantum, Pezze2016, Schnabel2016} is entering the stage where
ubiquitous quantum features of light and matter are being harnessed to deliver ultra-sensitive measuring devices for real-life applications
\cite{Schmidt2005, LIGO2011, Lucke2011, Taylor2013, Ockelon2013, Kovachy2015, Bouyer2016}.
Along with experimental advances, theoretical foundations for the field have been constantly developed. From the first proposals
of utilizing squeezed states in optical interferometry \cite{Caves1981}, through identification of fundamental  limits
in decoherence-free metrology \cite{Braunstein1992, Berry2000, Giovannetti2006}, general methods have been developed allowing to take into account the impact of realistic decoherence effects on the performance of metrological protocols \cite{Huelga1997, fujiwara2008fibre, Genoni2011, Escher2011, knysh2014true, demkowicz2012elusive, Demkowicz2013, jarzyna2017parameter} including the most general quantum adaptive strategies \cite{demkowicz2014using, sekatski2016quantum}.

Most of the available general methods is based on the integrated form of the dynamics of a quantum system represented mathematically as a quantum channel \cite{Escher2011, demkowicz2012elusive, demkowicz2014using}.
This poses a serious difficulty when the dynamics is provided in terms of a Master differential equation. In this case obtaining the analytical form of the integrated dynamics is often impossible. This fact significantly limits the
utility of the available methods making it often necessary to resort to numerical calculations instead of a more insightful analytical analysis.
This deficiency has been successfully addressed in case of single qubit dynamics, where the full description of performance of the most general quantum metrological protocols has been given \cite{sekatski2016quantum}. In particular, it has been shown that provided the noise is represented by a single Pauli operator which is not proportional to the Hamiltonian itself, one can apply an error correction procedure allowing to reach the Heisenberg-like, $T^2$,  scaling of Quantum Fisher Information (QFI), where $T$ is the total evolution time of the probe system. On the other hand, for all other kind of Markovian noise processes the optimal QFI is limited by a classical-like scaling bound proportional to $T$ and hence results in a standard $1/\sqrt{T}$ scaling  of precision.

This paper provides a general solution to the problem of determining  optimal performance of adaptive metrological schemes in a
unitary parameter estimation problem for arbitrary Markovian dynamics. We present an explicit recipe that allows to obtain the formulas for the
behavior of QFI $F_Q$ in the optimal metrological protocol, based solely on the operators appearing explicitly in the Master equation in the standard Gorini-Kossakowski-Lindblad-Sudarshan \cite{Breuer2002} form, with no need to integrate the dynamics whatsoever.
The probe dynamics we consider is given by
\begin{equation}
\label{eq:ME}
\frac{\t{d} \rho}{\t{d} t} =  - \text{i} \omega [H, \rho] + \sum_{j=1}^J L_j \rho L_j^\dagger -\frac{1}{2} \rho L_j^\dagger L_j - \frac{1}{2}L_j^\dagger L_j \rho,
\end{equation}
where $\omega$ is the frequency-like parameter to be estimated associated with the unitary dynamics generated by the Hamiltonian $H$,
 while $L_j$ are noise operators.
In particular, we show that if
\begin{align}
\label{eq:space}
\begin{split}
H \in \mathcal{S} = \t{span}_{\mathbb{R}}\{\openone, L_j^{\t{H}},  \text{i} L_j^{\t{AH}},
 &(L^\dagger_j L_{j^\prime})^{\t{H}}, \text{i} (L^\dagger_{j^\prime} L_{j})^{\t{AH}} \},
\end{split}
\end{align}
where $^{\t{H}}, ^{\t{AH}}$ denote the hermitian and the anti-hermitian part of an operator, then the QFI scales at most linearly with $T$, and the coefficient for the bound can be obtained from a solution of a simple semi-definite program. When restricted to a single qubit problem, this
condition is equivalent to the one given in \cite{sekatski2016quantum} requiring the noise
\emph{not} to be a single-rank Pauli linearly independent from the Hamiltonian.
If the above linear dependence condition is not satisfied
we discuss the possibility of implementing a ``quantum error-correction''-like protocol that yields
quadratic scaling of QFI in $T$.
 In the qubit case this is always possible \cite{sekatski2016quantum} using
 a scheme based on preparing a maximally
entangled state of the probe system and an equally dimensional ancilla.
 Here we demonstrate that in higher dimensions this is in general no longer the case, and this approach is not  always sufficient to overcome the effects of noise.

Further on we apply the newly developed quantitative methods to determine fundamental precision bounds in atomic metrological protocols involving
many-body interactions. This allows us in particular to derive for the first time fundamental precision bounds on non-linear metrological
 protocols in presence of decoherence. In absence of decoherence it is known that in the case of the $k$-body Hamiltonian non-linearity may
 help to improve the precision scaling of QFI to $T^2 N^{2k}$ where $N$ is the number of atoms involved \cite{Boixo2007,luis2007quantum, boixo2008quantum, choi2008bose, napolitano2010nonlinear, gross2010nonlinear, hall2012does, joo2012quantum, sewell2014ultrasensitive}.
 We show that in the case of the $k$-body Hamiltonian and $l$-body noise the linear dependence condition implies
 QFI to scale no better than $T N^{2k-l}$---a scaling formula identified in
 \cite{Beau2016} but only for GHZ states a and limited class of noise models.
 Apart from determining the scaling character of the bounds, we also provide explicit coefficients for the bounds in case of  linear and non-linear atomic interferometry in presence of single and two-body losses.
 Note that we focus here on unitary parameter estimation in presence of noise and do not analyze the problem of estimating the noise parameter itself. This last problem, while interesting, does not enjoy equally spectacular quantum gains thanks to the use of entangled states as the unitary parameter estimation case. Often a completely uncorrelated state proves to be optimal, as for example in the problem of estimating losses or dephasing strength, while in other cases entanglement between a single probe and a passive ancilla is sufficient to reach optimality
 \cite{Hotta2005, Ozawa2006, Kolodynski2013, Takeoka2016, Pirandola2017}


\section{Formulation of the problem}
Considering the Master equation given in Eq.~\eqref{eq:ME}, let us denote by  $\mathcal{E}^\omega_t$ the integrated form of the dynamics so that
\begin{equation}
\label{eq:integrated}
\rho^\omega_t = \mathcal{E}^\omega_t(\rho_0) = \sum_{j} K^{\omega}_{t,j} \rho_0 K_{t,j}^{\omega \dagger} (t),
\end{equation}
 where $K^{\omega}_{t,j}$ are Kraus operators of the evolution.

\begin{figure}[t]
\includegraphics[width=\columnwidth]{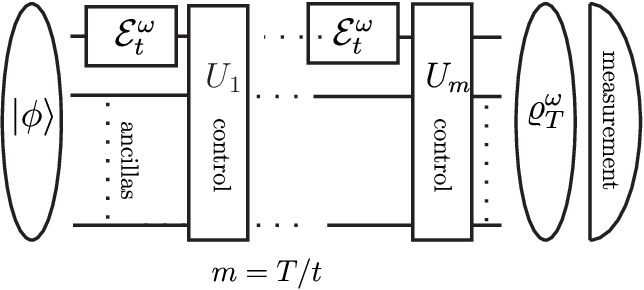}
\caption{General adaptive quantum metrological scheme. Total evolution time $T$ is divided into a number $m$ of $t$-long steps interleaved with general unitary controls. Collective measurement performed in the end allows to regard this scheme as a general adaptive protocol where measurement results at some stage of the protocol influence the control actions applied at later steps.
 This scheme may also mimic a parallel scheme where $m$ systems in an arbitrary initial entangled state
state are subject to evolution through $m$ parallel channels $\mathcal{E}^\omega_t$ for a time $t$.}
\label{fig:scheme}
\end{figure}

 The aim is to perform optimal estimation of $\omega$ parameter under the constraint of a fixed total evolution time $T$ under the most general adaptive quantum metrological scheme as depicted in Fig.~\ref{fig:scheme}.
Given the final state of the protocol $\varrho^\omega_T$, the fundamental limitation on the precision of estimating $\omega$
is given in terms of quantum Cram{\'e}r-Rao bound:
\begin{equation}
\label{eq:qfi}
\Delta\omega \geq \frac{1}{\sqrt{F_Q}}, \ F_Q = 2 \sum_{a b} \frac{|\bra{a}\dot{\varrho}^\omega_T|\ket{b}|^2}{\lambda_a + \lambda_b} \
\end{equation}
where $F_Q$  is QFI for unitary encoding, dot signifies $\frac{d}{d\omega}$, while $\ket{a}$, $\lambda_a$ are eigevectors and eigenvalues of $\varrho^\omega_T$. In what follows we will use QFI as the figure of merit. Direct maximization of  QFI of the final state over all tunable elements in the protocol, i.e. input state and controls is a virtually impossible task unless a decoherence-free case is considered where adaptiveness is useless \cite{Giovannetti2006}.

Luckly, provided the integrated form of the dynamics in the form of \eqref{eq:integrated} is given, one can apply the methods from \cite{demkowicz2014using, sekatski2016quantum} that allow to obtain a universally valid upper bound on QFI valid for arbitrary adaptive strategy, and hence a lower bound on uncertainty. The bound utilizes the observation that given a quantum evolution in the form of Kraus representation \eqref{eq:integrated} one can consider equivalent Kraus representation $\tilde{K}^{t,j}= \sum_{j^\prime} u^{\omega}_{jj^\prime}K^{\omega}_{t,j^\prime}$ consisting of operators related by a unitary matrix with the original ones, or written in a more concise notation
$\tilde{\mathbf{K}} =  u \mathbf{K}$, where $\tilde{\mathbf{K}} = [\tilde{K}^{\omega}_{t,0},\tilde{K}^{\omega}_{t,1},\dots]^T$.

 The bound on QFI for the $m=T/t$ step adaptive protocol  is then given in terms of minimization over different Kraus representations
 and reads:
\begin{align}
\begin{split}
F_Q \leq 4 \min_{\tilde{\mathbf{K}}_{\omega},x} & \left\lbrace   m \lVert \alpha \rVert \right.  \\
& \left.  + m(m-1) \lVert \beta \rVert(x \lVert \alpha \rVert + \lVert \beta \rVert + 1/x) \right\rbrace ,
\end{split}\label{eq:qfibound}
\end{align}
where
\begin{equation}
\alpha :=  \dot{\tilde{\mathbf{K}}}^{\dagger} \dot{\tilde{\mathbf{K}}}, \quad \beta := -\text{i} \dot{\tilde{\mathbf{K}}}^{\dagger} \tilde{\mathbf{K}},
\end{equation}
 $\lVert . \rVert$ is the operator norm and $x$ is a positive real parameter minimization over which helps to further tighten the bound.
The crucial element here is that the unitary matrix $u$ can explicitly depend on the estimated parameter $\omega$. Using the fact that $u^\dag \dot u = -\ii h$ for some hermitian matrix $h$, one notes that this dependence enters the computation of $\alpha$ and $\beta$ via the substitution:
\begin{equation}
\dot{\tilde{\mathbf{K}}}=\dot{\mathbf{K}} -\text{i}  h  \mathbf{K},
\quad {\tilde{\mathbf{K}}} = {{\mathbf{K}}}.
\end{equation}

In the most general adaptive approach it is always advantageous to take the limit $t \rightarrow 0$ as in this case  the use of adaptive gates potentially provides the greatest benefits.  Note that the limit is taken only in the duration of the sensing map $\mathcal{E}_t^\omega$ and it does not influence the duration of the intermediate unitary control gates. In fact in our model we assume that the time needed to perform control gates is not included in the total resource count and hence the continuous limit $t \rightarrow 0$ does not affect the control gates time. At the same time this limit allows us to have arbitrary many control gates over the course of the sensing process and hence results in the most general adaptive strategy.

In order to keep this in mind, from now on we will therefore replace $t$ with $dt$. Taking this limit we may now utilize known relations between noise operators $L_j$ and the corresponding Kraus operators $K_j$ in order to write an explicit Kraus representation for the dynamics in the lowest order in $dt$:
\begin{align}
\label{eq:kraus}
K_0 &= \openone - \left(\frac{1}{2} \mathbf{L}^\dagger \mathbf{L} + \text{i} H \omega \right) dt + O(dt^{2}), \\
K_j &= L_j \sqrt{dt} + O(dt^{\frac{3}{2}}) \  (j=1,\dots,J)
\end{align}
where $\mathbf{L} = [L_1,L_2,\dots]^T$. Note that Kraus operators' index starts from $0$ while the noise opeartors'
index starts from $1$.
This approximation correctly recovers the dynamics in the linear order in $dt$ and captures all the features of Markovian evolution.

Because of the different $dt$ scaling appearing in $K_0$ and $K_{j\geq1}$ operators it will be convenient to introduce the following structure
of the matrix $h$:
\begin{equation}
h= \mat{c|ccc}
{h_{00} & &\mathbf{h}^\dagger & \\ \hline &&& \\  \mathbf{h} & &\mathfrak{h} & \\
& & &}.
\end{equation}
Minimization over different Kraus representations in  \eqref{eq:qfibound} amounts now to minimization over the matrix $h$ and  taking into account that we consider the  limit $dt\rightarrow 0$ we get
\begin{align}
\label{eq:qfiboundt}
\begin{split}
F_Q \leq 4 \min_{h,x} & \left\lbrace   T  \lVert \alpha \rVert dt^{-1} \right.\\
& \left.+ T^2 \lVert \beta \rVert dt^{-2}(x \lVert \alpha \rVert + \lVert \beta \rVert + 1/x)\right\rbrace .
\end{split}
\end{align}
The most interesting challenge posed be the  above formula is to determine the situation where $F_Q$ is limited by a bound scaling linearly in $T$ or
where the bound scales as $T^2$, in which case achieving the Heisenberg scaling may be possible.

\section{$H \in \mathcal{S}$: $T$ scaling of QFI}
\label{sec:linear}
The bound will scale linearly in $T$ if we are able to find $h$ which makes $\beta=0 + O(dt^2)$ as well
as $\alpha = \alpha^{(1)} dt + O(dt^2)$, since then by choosing $x=1/\sqrt{dt}$ we will get
\begin{equation}
\label{eq:alphabound}
F_Q \leq 4 \lVert \alpha^{(1)}\rVert T
\end{equation}
in the limit $dt \rightarrow 0$.
Let us write explicitly $\beta$ in terms of $\mathbf{L}$, $h$, $H$  in leading orders in $dt$:
\begin{align}
\begin{split}
\beta = & H dt +  h_{00}(\openone - \mathbf{L}^\dagger \mathbf{L} dt) \\
&+ (\mathbf{h}^\dagger \mathbf{L}+\mathbf{L}^\dagger \mathbf{h})\sqrt{dt} + \mathbf{L}^\dagger \mathfrak{h} \mathbf{L} dt  + O(dt^{\frac{3}{2}}).
\end{split}
\end{align}
Let us denote time expansion coefficients of $h$ as follows: $h = \sum_{k=0,\frac{1}{2},1,\dots} h^{(k)} dt^{k}$.
We now investigate the condition $\beta=0$ order by order in time. Making $\beta=0$ in orders $O(dt^0)$ and $O(dt^{\frac{1}{2}})$ implies
 that  $h_{00}^{(0)}= h_{00}^{(\frac{1}{2})} = 0$ as well as $\mathbf{h}^{(0)} = 0$.  The first non-trivial condition appears in $O(dt)$ order as this is the order where the Hamiltonian $H$
contributes and setting $h=0$ is not enough to get $\beta^{(1)}=0$.
 With the above substitutions we may write the linear order coefficient of $\beta$:
 \begin{equation}
 \label{eq:beta}
 \beta^{(1)} = H  + {h}_{00}^{(1)} \openone +  {\mathbf{h}}^{\dagger (\frac{1}{2})} \mathbf{L} +\mathbf{L}^\dagger \mathbf{h}^{({\frac{1}{2}})} +
  \mathbf{L}^\dagger \mathfrak{h}^{(0)} \mathbf{L}.
 \end{equation}
Taking into account that $h$ is hermitian, this coefficient can be made zero if and only if the Hamiltonian
$H \in \mathcal{S}$,
where subspace $\mathcal{S}$ is defined in \eqref{eq:space}.
Analysing the next order
we get:
\begin{equation}
\beta^{(\frac{3}{2})}  =  h_{00}^{(\frac{3}{2})}\openone +  \mathbf{h}^{(1)\dagger } \mathbf{L} +\mathbf{L}^\dagger \mathbf{h}^{(1)} +
  \mathbf{L}^\dagger \mathfrak{h}^{(1/2)} \mathbf{L}.
\end{equation}
We see that none of the coefficients appearing here appeared before when considering $\beta^{(1)}$ so we may put them all equal zero guaranteeing that $\beta = 0 + O(t^2)$, and proving the linear scaling of QFI.

In order to obtain a quantitative bound, as
given in Eq.~\eqref{eq:alphabound}, let us now focus on the operator $\alpha$.
Taking into account that $h_{00}^{(0)}= h_{00}^{(\frac{1}{2})} = 0$ as well as $\mathbf{h}^{(0)} = 0$, the first nontrivial order is linear in $dt$ and the corresponding coefficient reads:
\begin{equation}
\label{eq:alpha}
\alpha^{(1)} = \left(\mathbf{h}^{(\frac{1}{2})} \openone + \mathfrak{h}^{(0)}\mathbf{L} \right)^\dagger \left( \mathbf{h}^{(\frac{1}{2})} \openone + \mathfrak{h}^{(0)}\mathbf{L}\right ).
\end{equation}
We now need to minimize the operator norm of the above coefficient over $h$ in order to get the tightest bound:
\begin{align}
\begin{split}
F_Q \leq  4 T & \min_{\{h_{00}^{(1)},\mathbf{h}^{(\frac{1}{2})},\mathfrak{h}^{(0)}\}}  \lVert \alpha^{(1)} \rVert , \\
 &\textrm{subject to: } \beta^{(1)}=0.
\end{split}
\end{align}
Only in some particular cases this can be done analytically. Fortunately, the above problem can be implemented as a semi-definite program, as described explicitly in Appendix \ref{sec:sdp}.
The implementation is similar to the one presented in \cite{demkowicz2012elusive} for the Kraus operator formulation.

Since the bound \eqref{eq:alphabound} involves the operator norm it may not be immediate to apply
it in infinite dimensional cases where the operators appearing in the Master equation are unbounded.
This is for example the case when dealing with continuous variable systems.
Following the original derivation of the bound~\eqref{eq:qfibound}, however,
one can refine it by taking into account some properties of the state
utilized in the protocol. In particular it might be that
the states we deal with are restricted to consist of finite number of photons/atoms, or
at least have the mean number of particles fixed.
It might also, be the case that by various super-selection rules
the whole Hilbert space is not available and the bound can be tightened by analysing the operator norm of $\alpha^{(1)}$ separately in different super-selection sectors.
This will amount to calculation of operator norms on subspaces. Moreover,
provided sufficient information on the time evolution of the probe state is given, the
bound may even be formulated as an integral over interrogation time $T$ of a time-dependent quantity.
Namely
\be
F_Q\leq 4 \int_0^T \min \, \langle \alpha^{(1)} \rangle_t dt, \quad \textrm{subject to:  }\beta^{(1)}=0,
\ee
with $\langle \alpha^{(1)} \rangle_t= \tr \rho_t \alpha^{(1)}$ setting a limit on the gain in QFI at a given instant of time, and $\rho_t$ is the state of the system at time $t$ --- see Appendix \ref{sec:constraints} for the details how to tighten the bound in these cases.

In what follows, it will be convenient to assume that the Master equation \eqref{eq:ME} is given in the so called canonical form
\cite{Pearle2012, hall2014canonical}, where all noise operators are traceless and orthogonal $\tr L_j= \tr L_k^\dag L_j=0$, see also \ref{sec:noise operators}.

Let us now briefly comment on the single qubit case discussed in \cite{sekatski2016quantum}.
Since $L_i$ operators in the canonical form are traceless there can be written as complex combinations of Pauli matrices.
The condition required to get a better than linear scaling of QFI discussed there was that
 the noise is a single-rank Pauli linearly independent from the Hamiltonian.
Mathematically this means that there is only one $L_j \propto \sigma_{\vec{n}} \not\propto H$.
Note that if we had two linearly independent $L_j$,
they would lead to $\mathcal{S}$ being  the full space of $2 \times 2$  hermitian matrices, thanks to the fact that products $L_j^\dagger L_{j^\prime}$ appear in the definition of $\mathcal{S}$.
Moreover, even with a single $L_i$ which is not proportional to a hermitian matrix,
we would have two linear independent hermitian matrices from its hermitian and anti-hermitian part and
hence again the generated $\mathcal{S}$ would be equal to the whole space of hermitian matrices.
Consequently, in the qubit case, the $H \in \mathcal{S}$ condition is
equivalent to the one discussed in \cite{sekatski2016quantum}.

\section{$H \notin \mathcal{S}$: Possibility of $T^2$ scaling of QFI}
\label{sec:quadratic}

If $H \notin \mathcal {S}$ and hence $\beta^{(1)}$ cannot be made zero
then the second term in the bound~\eqref{eq:qfiboundt} will not vanish (due to $\lVert \beta \rVert^2$ scaling as $dt^2$ and canceling with  $1/dt^2$ term) and will result in an upper bound scaling as $T^2$. This gives hopes for the Heisenberg scaling of precision. Below we discuss the possibility to construct an adaptive quantum error correction inspired strategy that allows to achieve a $T^2$ scaling of QFI and discuss some concrete examples.

Aside the probe system we allow for an additional ancillary system  on which the evolution acts trivially. Let $\varrho =\proj{\phi}$ denote the input probe-ancilla state.
 The adaptive protocol we consider consists of
 intertwining of infinitesimal-time probe evolution
 \be
 \mathcal{E}^{\omega}_{dt}(\varrho) = \varrho + \Big(- \text{i} \omega [H,\varrho] + \mathcal{L}(\varrho)\Big)dt + O(dt^2),
 \ee
 where $\mathcal{L}$ represents the noise part of the Master equation \eqref{eq:ME}, and the error correction map $C$ applied after each infinitesimal time step $dt$.
 Hence, the final state of the probe and ancilla systems after the total evolution time $T$, i.e.
 after $T/dt$ applications of the evolution-correction step, reads:
 \be
 \varrho_{T}^\omega =\mathcal{C}_T^\omega (\varrho) =  (C\circ \cE^\omega_{dt})^{\circ\frac{T}{dt}}(\varrho).
 \ee
 Formally, in the above formula we should write $\cE^\omega_{dt} \otimes \mathcal{I}$  instead of
 $\cE^\omega_{dt}$, as the map acts also on the ancillary system in a trivial way.
 From now on, for simplicity of notation, we will assume that whenever an operator defined on the probe system alone acts on the extended probe-ancilla system it should be understood as extended in a trivial way.

 To simplify the exposition we assume that estimation of $\omega$ is made around $\omega_0=0$ point. Otherwise by means of active control one can always compensate for the nonzero rotation term  $-\ii \omega_0[H,\rho]$ in the master equation---note that in this case the error correction operation $C$ may in general depend on $\omega_0$.
 Since QFI depends on the state and its first derivative at $\omega_0$, see Eq.~\eqref{eq:qfi},
 it is enough to consider the first order expansion of the final state in the estimated parameter $\omega$: $  \varrho_T^\omega= \varrho_T^0 +\omega\, \dot{\varrho}_T^{0} +O(\omega^2)$.
 The zeroth order $\varrho_T^0 = \mathcal{C}_T^0(\varrho)$ is simply the action on the input state
 of the dynamics where the Hamiltonian part is dropped, while the first order term
   \be
    \dot{\varrho}_T^0 = - \ii \int_0^T \mathcal{C}_{T-t}^0
   \Big(\big[H, \mathcal{C}_{t}^0(\varrho)\big] \Big) dt
 \label{eq:gen1st}
 \ee
involves terms of $\mathcal{C}_T^0$ where the  Hamiltonian part of the dynamics enters linearly at different times.

Our goal is to design a protocol that protects the system from decoherence in a way that its QFI grows quadratically in $T$ and hence mimics the performance of noiseless frequency estimation.
First of all, we demand that our protocol preserves the initial state in absence of the unitary evolution $\varrho_t^0 = \mathcal{C}_t^0(\varrho) = \ket{\phi}\bra{\phi}$.
Furthermore, let us define a decoherence-free qubit subspace $\cH_Q=\{\ket{\phi},\ket{\xi} \}$, spanned by the input state
as well as an orthogonal state $\ket{\xi}$ such that $C([H, \ket{\phi}\bra{\phi}]) = c (\ket{\xi}\bra{\phi} - \ket{\phi}\bra{\xi})$,
 $c \in \mathbb{R}$. The error-correction step $C$ projects the state after being acted on with $H$ back onto the subspace $\cH_Q$.
We also assume that $c \neq 0$ since otherwise the parameter $\omega$ would not be imprinted on the state at all.
This allows us to simplify the the expression in Eq.~\eqref{eq:gen1st}
 \begin{align}
 \dot{\varrho}_{T}^0 &= \ii  c \int_0^T \mathcal{C}_{T-t}^0 \Big(\coh{\phi}{\xi}-\coh{\xi}{\phi}\Big) dt.
 \label{eq:1st}
 \end{align}
 In addition, we require that the control-assisted evolution $\mathcal{C}_{T-t}^0$ preserves the coherence
  $\coh{\phi}{\xi}$, which leads to
  $\dot{\varrho}_{T}^0 = \ii \,c\, T \big(\coh{\phi}{\xi}-\coh{\xi}{\phi}\big)$.
  Calculating QFI with eigenvectors $\{\ket{\phi},\ket{\xi}\}$ and $\lambda_{\phi}=1$ yields quadratic $F_Q = 4 c^2 T^2$ as in the case of noiseless frequency estimation.
  Otherwise, had the evolution damped the coherence term, resulting in  $||\mathcal{C}_{t}^0\Big(\coh{\phi}{\xi}-\coh{\xi}{\phi}\Big) ||\leq e^{-\lambda t}$ this would not be possible.  Hence in summary, the requirement for the error correction map $C$ are the following:
\begin{align}
\t{(i)}\quad C\big(\proj{\phi} + dt \mathcal{L}(\proj{\phi})\big)&=\proj{\phi} + O(dt^2)\\
\t{(ii)}\quad C\big(\coh{\xi}{\phi} + dt \mathcal{L}(\coh{\xi}{\phi})\big)&=\coh{\xi}{\phi}+O(dt^2).
\end{align}

In fact, the linearity and the trace preserving properties of $C$, together with the above conditions, imply
 $C\big(\proj{\xi} + dt \mathcal{L}(\proj{\xi})\big)=\proj{\xi} + O(dt^2)$.
 As a result, the requirement for QFI to grow quadratically amounts to the requirement of existence of a  two-dimensional subspace protected from decoherence up to the linear order in time.
We may therefore utilize known results from approximate quantum error correction
literature, which in this case reduce to the standard error-correction relation \cite{Knill1997}
for the set of error operators consisting of $L_i$ and the identity operator \cite{Cedric2011}:
\begin{align}
\label{eq:ecconditions}
&\t{(a)}\quad\bra{\phi} H\ket{\xi}\neq 0,\\
&\t{(b)}\quad\bra{\phi} L_k^\dag L_j \ket{\xi}=\bra{\phi} L_j \ket{\xi}=0,
\label{eq:conditions2}\\
&\t{(c)}\quad\bra{\phi}L_k^\dag L_j \ket{\phi}= \bra{\xi}L_k^\dag L_j \ket{\xi},
\label{eq:conditions1}
\end{align}
for all $k$ and $j$, where  (a) is an additional requirement that needs to be satisfied in order to keep non-trivial  unitary evolution in the qubit subspace $\cH_Q$. Step by step derivation of the above conditions is provided in Appendix \ref{sec:errorcorrection}.

Following the way the single-qubit error-correction protocols were applied in quantum metrology \cite{Arrad2014, Kessler2014a, Dur2014, sekatski2016quantum}
 a natural choice for  $\ket{\phi}$ is the maximally entangled state of probe+ancilla
 $\ket{\phi}=\frac{1}{\sqrt{d}}\sum_{i} \ket{i}\otimes \ket{i}$.
 Recall that in the canonical form of Master equation all $\mathcal{L}$ are traceless and orthogonal $\tr L_j= \tr L_k^\dag L_j=0$, see \ref{sec:noise operators}. Hilbert-Schmidt orthogonality of $L_j$ is automatically transferred to
orthogonality of $L_j \ket{\phi}$ vectors (where $L_j$ should be in fact understood here as $L_j \otimes \openone$).
We then decompose the Hamiltonian $H = H_\perp + H_\parallel$ such that $H_\parallel \in \mathcal{S}$ while nonzero
$H_\perp \in \mathcal{S}_\perp$ is orthogonal to all operators in $\mathcal{S}$. If we  now take $\ket{\xi} =\frac{H_\perp \ket{\phi}}{\lVert H_\perp\ket{\phi}\rVert}$
(note it is by construction orthogonal to $\ket{\phi}$ as $H_\perp$ is in particular orthogonal to the identity operator), then one automatically satisfies the first two conditions. Condition (a) follows from $\bra{\phi}H_\perp  H \ket{\phi}\propto \tr H_\perp H \neq 0$, while (b) follows from
\begin{align}
\begin{split}
&\bra{\phi} L_k^\dag  L_j H_\perp \ket{\phi}\propto \tr H_\perp  L_k^\dag L_j=0, \\
& \bra{\phi}L_j H_\perp \ket{\phi} = 0,
\end{split}
\end{align}
as $H_\perp$ is orthogonal to $\mathcal{S}$ with respect to the Hilbert-Schmidt scalar product.
For the qubit case \cite{sekatski2016quantum} this construction also guarantees condition c) to be satisfied, as $H \notin \mathcal{S}$ implies only one $L_i=L$.
Let $\ket{0}$, $\ket{1}$ be the eigenbasis of $H_\perp$:
    $H_\perp \ket{i}= \lambda (-1)^i \ket{i}$ ($H_\perp$ is orthogonal to $\openone$ and hence has $\pm \lambda$ eigenvalues). As a result   $\ket{\xi} \propto H_\perp \ket{\phi} \propto (\ket{0} \otimes\ket{0} - \ket{1}\otimes\ket{1})$, and consequently:  $\bra{\phi}L^\dagger L \ket{\phi}= \bra{\xi}L^\dagger L \ket{\xi} = \t{Tr}(L^\dag L)/2$.
In this case one can state that Eq.~\eqref{eq:space} is an if and only if condition for impossibility of getting QFI scaling quadratically with $T$.


In higher dimensions, however, an error-correction scheme based on the use of the maximally entangled state of system+ancilla will not work in general---see the Note Added at the end of the Conclusions section with a reference to the paper \cite{zhou2018achieving}, where a universal construction of a quantum error-correction protocol satisfying all the required conditions has been provided whenever $H \notin\mathcal{S}$. This also, shows that the conditions which in our approach could be regarded as sufficient for quadratic scaling are actually also necessary.

\section{Atomic interferometry with one and two-body losses}
\label{sec:losses}
We will now demonstrate an application of the developed methods to provide bounds in atomic interferometry models
where two-body effects can be placed both in the noise part (two-body losses) or in the Hamiltonian part (the non-linear metrology model).

Let us consider a Bose-Einstein Condensate (BEC) system of two level atoms, where dynamics is described by the following master equation:
\begin{equation}
\frac{\t{d} \rho}{\t{d} t} =  - \text{i} \omega \left[H^{(k)}, \rho\right] + \mathcal{L}^{(1)}(\rho) + \mathcal{L}^{(2)}(\rho),
\end{equation}
where
\begin{align}
\mathcal{L}^{(1)}(\rho) &= \sum_{i=1}^2 \gamma_{i}\left(a_i \rho a_i^\dagger - \frac{1}{2}\left\{a_i^\dagger a_i, \rho  \right\} \right), \\
\begin{split}
\mathcal{L}^{(2)}(\rho) &= \sum_{i=1}^2 \gamma_{ii}\left(a_i^2\rho a_i^{\dagger 2} - \frac{1}{2}\left\{a_i^{\dagger 2}a_i^2, \rho  \right\} \right)  \\
& + \gamma_{12}\left(a_1a_2\rho a_1^{\dagger}a_2^{\dagger} - \frac{1}{2}\left\{a_1^{\dagger}a_2^{\dagger}a_1a_2, \rho  \right\} \right),
\end{split}
\end{align}
represent one-body and two-body loss processes with respective loss coefficients $\gamma_i, \gamma_{ij}$ and
$a_i$ representing anihilation operators removing an atom from the $i$-th mode. The corresponding noise operators
read: $L^{(1)}_i = \sqrt{\gamma_i} a_i$, $L_{ij}^{(2)}= \sqrt{\gamma_{ij}}a_i a_j$.
We will consider two different Hamiltonians that are associated with the sensing part of the dynamics
\begin{align}
H^{(k=1)} &  =\frac{1}{2}(a_1^\dagger a_1 -a_2^\dagger a_2), \\
H^{(k=2)} &  =\frac{1}{4}:(a_1^\dagger a_1 -a_2^\dagger a_2)^2:,
 \end{align}
which correspond to linear and non-linear metrological scenarios.
For the clarity of presentation, we have put the normal ordering operation in the definition of  $H^{(2)}$ in order to make sure that
we take into account only terms that appear due to interaction between two different particles.

Let us start with the linear Hamiltonian  case $k=1$, but keep both the single and two-body losses processes. This kind of model is
well tailored to analyze matrological BEC experiments such as e.g. magnetometry experiment using spin-squeezed BEC  \cite{Ockelon2013}.
Let us calculate $\beta^{(1)}$ according to Eq~.\eqref{eq:beta}:
\begin{multline}
\label{eq:losses1}
\beta^{(1)} = \frac{1}{2}(a_1^\dagger a_1 - a_2^\dagger a_2) + h^{(1)}_{00}\openone + \left(\sum_i \sqrt{\gamma_i} h^{(\frac{1}{2})}_i a_i  + h.c.\right)\\
+ \sum_{ij} \mathfrak{h}^{(0)}_{ij} \sqrt{\gamma_i \gamma_j} a_i^\dagger a_j + \dots .
\end{multline}
We now ask about the possibility of choosing entries of $h$ in a way to set the above quantity to zero. Note that we have not written explicitly the noise operators related to two-body losses. The reason for that is that operators related to two-body losses appearing in the above equation would be of the form
$a_i a_j, a_i^{\dagger} a_j, a^\dagger_i a^\dagger_j a_i^\prime a_j^\prime$ and would be linearly independent of the operators appearing in the Hamiltonian part. Hence trying to set $\beta^{(1)}=0$ we need to focus on one-body losses operators only. We are free to put
all coefficients of $h$ in front of terms related with two-body losses equal to zero. If we succeed in setting $\beta^{(1)}=0$ using only one-body losses operators this will also imply that two-body losses are irrelevant in trying to assess the fundamental precision limit on frequency estimation in this case. By inspecting Eq.\eqref{eq:losses1} it is clear that we can make $\beta^{(1)}=0$ by choosing
$h^{(1)}_{00}=0$, $h^{(\frac{1}{2})}_i=0$, $\mathfrak{h}^{(0)}_{11}=
-\frac{1}{2} \gamma_1^{-1}$,  $\mathfrak{h}_{22}^{(0)}=\frac{1}{2} \gamma_2^{-1}$, $\mathfrak{h}_{12}^{(0)}=0$.

We should remember, however, that when deriving the final bound using Eq.~\eqref{eq:alpha} we face the problem that operators appearing under the operator norm are unbounded and hence the bound formally will be infinite and hence useless.
Physically this is due to the fact that we have not set any constraints on the number of atoms we use in the experiment.
From now on we will assume we have  $N$ atoms at our disposal and at every adaptive step
we replace the lost atoms with fresh ones keeping the number of atoms constant.
We discuss this approach in detail in Appendix \ref{sec:constraints} and argue that by doing so we do note lose generality of our bounds.
Thanks to this, we are able to write  $a_1^\dagger a_1 + a_2^\dagger a_2 = N \openone$ when calculating the operator norm of $\alpha^{(1)}$, remembering that in the end we operate in a fixed particle number-subspace.

Let us now go back to Eq.~\eqref{eq:losses1}. With fixed particle number constraint imposed, $a_1^\dagger a_1$, $a_2^\dagger a_2$ and $\openone$ are no longer independent operators. This gives us an additional freedom in choosing coefficients of $h$ in order to keep $\beta^{(1)}=0$,  namely we
can take $h_{00}^{(1)} = -N\xi$ ,  $\mathfrak{h}^{(0)}_{11}=
\gamma_1^{-1}(\xi-\frac{1}{2})$,  $\mathfrak{h}^{(0)}_{22}= \gamma_2^{-1}(\xi + \frac{1}{2})$, with $\xi$ being a free parameter.
The QFI bound can be now obtained by minimizing $\lVert \alpha \lVert$ over $\xi$:
\begin{equation}
F_Q \leq T  \min_\xi \lVert \gamma_1^{-1}(2\xi-1)^2 a_1^\dagger a_1 +  \gamma_2^{-1}(2\xi+1)^2 a_2^\dagger a_2 \lVert.
\end{equation}
Operators $a_1^\dagger a_1$, $a_2^\dagger a_2$ commute and their common basis is $\ket{n,N-n}$ (where $n$ is the number of atoms in mode $1$) hence in the above minimization we can replace $a_1^\dagger a_1$ with $n$ and $a_2^\dagger a_2$ with $N-n$:
\begin{align}
\begin{split}
F_Q \leq  T \min_\xi \max_{0 \leq n \leq N}& \gamma_1^{-1}(2\xi-1)^2 n \\
&+  \gamma_2^{-1}(2\xi+1)^2(N-n).
\end{split}
\end{align}
The minimum is achieved for $\xi$ that satisfies $\gamma_1^{-1}(2\xi-1)^2 = \gamma_2^{-1}(2\xi+1)^2$ and reads:
\begin{equation}
\label{eq:bound1losses}
F_Q \leq \frac{4 T  N}{(\sqrt{\gamma_1} + \sqrt{\gamma_2})^2}.
\end{equation}
This bound indeed agrees with a known bound of $N$ particle interferometry with losses \cite{Kolodynski2010, Knysh2010, Escher2011, demkowicz2012elusive}
\begin{equation}
F_Q \leq \frac{4 T t N}{\left(\sqrt{\frac{1-\eta_1}{\eta_1}} + \sqrt{\frac{1-\eta_2}{\eta_2}} \right)^2},
\end{equation}
where $\eta_i = e^{-\gamma_i t}$ after taking the limit  $t \rightarrow 0$.
Note, however, that the derivation presented in this paper did not require any educated guess
\cite{Escher2011} nor numerically
indicated optimal form of Kraus representation \cite{demkowicz2012elusive},
but resulted in a purely algebraic analysis of the noise operators and the Hamiltonian appearing in the Master equation. Moreover, when deriving the bound we could clearly see that two-body losses
do not have an impact on the bound.

We now move on to study the fundamental bounds in a non-linear metrological model with $k=2$,
in which case
\begin{align}
\begin{split}
\beta^{(1)} = &\frac{1}{4}(a_1^{\dagger 2}a_1^2 + a_2^{\dagger 2}a_2^2 - 2 a_1^\dagger a_2^\dagger a_1 a_2) + h^{(1)}_{00} \openone \\
&+\mathfrak{h}^{(0)}_{11,11} \gamma_{11} a_1^{\dagger 2}a_1^2 + \mathfrak{h}^{(0)}_{22,22} \gamma_{22}  a_2^{\dagger 2}a_2^2 \\
&+\mathfrak{h}^{(0)}_{12,12} \gamma_{12} a_1^\dagger a_2 \dagger a_1 a_2 + \dots,
\end{split}
\end{align}
where we explicitly wrote only terms relevant for further discussion. In particular, we can ignore the one-body loss operators as they are linearly independent from the Hamiltonian and hence will not contribute to the bound.
Similarly as in the linear case, we assume we deal with $N$-atom states. Hence, we will
utilize the fact that $(a_1^\dagger a_1 + a_2^\dagger a_2)^2 = N^2 \openone$ implying the following relation:
$a_1^{\dagger 2} a_1^2 + a_2^{\dagger 2}a_2^2 + 2 a_1^\dagger a_2^\dagger a_1 a_2 = N(N-1)\openone$, which allows us
to introduce again a free parameter $\xi$ into coefficients of $h$:
$\mathfrak{h}^{(0)}_{11,11} = \frac{1}{4}\gamma_{11}^{-1}(\xi -1)$, $\mathfrak{h}^{(0)}_{22,22} = \frac{1}{4}\gamma_{22}^{-1}(\xi -1)$,
$\mathfrak{h}^{(0)}_{12,12} = \frac{1}{2}\gamma_{12}^{-1}(\xi +1)$, $h_{00}^{(1)} = -\frac{1}{4}\xi N(N-1)$.
Using Eq.~\eqref{eq:qfibound} we arrive at the following bound:
\begin{align}
\label{eq:bound2lossesder}
\begin{split}
F_Q \leq & \min_\xi  \frac{T}{4}\bigg\lVert2 \gamma_{12}^{-1}(\xi+1)^2 N(N-1)\openone  \\
 &+ a_1^{\dagger 2} a_1^2\left(\gamma_{11}^{-1}(\xi-1)^2 - 2 \gamma_{12}^{-1}(\xi+1)^2 \right)  \\
& + a_2^{\dagger 2} a_2^2\left(\gamma_{22}^{-1}(\xi-1)^2 - 2 \gamma_{12}^{-1}(\xi+1)^2 \right) \bigg\lVert.
 \end{split}
\end{align}
Since all operators under the operator norm commute and have common eigenbasis $\ket{n,N-n}$, we can write the above bound explicitly
replacing $a_1^{\dagger 2}a_1^2$ with $n(n-1)$ and $a_2^{\dagger 2}a_2^2$ with $(N-n)(N-n-1)$. Calculating the operator norm amounts now to maximization over $n$. In case $\gamma_{11}=\gamma_{22}$ the above problem has an explicit solution which in the limit of large $N$ reads:
\begin{equation}
\label{eq:bound2losses}
F_Q \leq \frac{N^2 T }{\gamma_{12}} \begin{cases}\frac{2}{(1+\sqrt{\lambda})^2},  & \lambda  \geq 1  \\ \frac{1}{1+\lambda}, & \lambda < 1 \end{cases}.
\end{equation}
where $\lambda = 2 \gamma_{11}/\gamma_{12}$. To the best of our knowledge this is the first example of a fundamental bound in the non-linear metrology model taking into account many-body decoherence effects.
Details of the above derivation as well as the discussion of the case $\gamma_{11} \neq \gamma_{22}$ is presented in Appendix \ref{sec:twobody}.


\section{Quantum metrology with general many-body interactions}
Let us now investigate what can be said in general concerning the
fundamental bounds in non-linear metrology models with many-body interactions,
without specifying the actual form of the dynamics but just its non-linear character and trying to identify the resulting characteristic scaling of QFI.

Consider a system of $N$ atoms, where the hamiltonian part is a result of $k-body$ interactions while
the noise part is of $l-body$ type. To be more specific we consider the dynamics of the form:
\begin{align}
\label{eq:nonlinear}
\begin{split}
&\frac{\t{d} \rho}{\t{d} t} =  - \text{i} \omega \left[\sum_{\nu \in \combs_k}H_{\nu}, \rho\right]  \\
&+\gamma \sum_{\mu \in \combs_l, j} L_{\mu,j} \rho L_{\mu,j}^\dagger -\frac{1}{2} \rho L_{\mu,j}^\dagger L_{\mu,j} - \frac{1}{2}L_{\mu,j}^\dagger L_{\mu,j} \rho,
\end{split}
\end{align}
where $\combs_k = \{(i_1,\dots,i_k)\}$ represents all $k$ element combinations of the $N$ element set, and the
operator index $\nu \in \combs_k$ denotes particles that a given operator acts on. We have also introduced a positive coefficient $\gamma$ in order to be able to discuss effects of rescaling the noise strength. In what follows we will assume that
$k,l \ll N$.
This is a general scenario considered in the field of
non-linear metrology \cite{Boixo2007,luis2007quantum, boixo2008quantum, choi2008bose, napolitano2010nonlinear, gross2010nonlinear, hall2012does, joo2012quantum, sewell2014ultrasensitive}, but very often without the noise part.
Including the noise part in this form in the analysis is extremely challenging and has only been analyzed for a very specific noise models and only in case of input GHZ states \cite{Beau2016}.

\begin{figure}[t]
\includegraphics[width=\columnwidth]{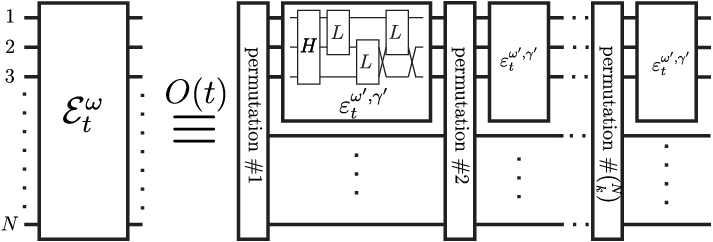}
\caption{Equivalent (up to linear order in $t$) representation of the $N$ particle dynamics in the form of a subsequent action
of the $k$-particle Hamiltonian $H$ and $l$-particle noise $L$ on all $n$ particle subsets of the total $N$ particles
(the circuit is given for $n = k = 3\geq l = 2$).
Since the number of applications of the noise part is here enhanced by a factor $\bi_l  \propto N^{n-l}$,
we need to rescale the noise coefficient in the above scheme to $\gamma^\prime  \propto   \gamma/\bi_l$  to preserve the equivalence.
 Similarly if  $n>k$ we would need to rescale the frequency parameter to
$\omega^\prime = \omega/\bi_k$. This representation allows  to calculate the bound on QFI for the whole dynamics by analysing the properties of an elementary $n$-particle subchannel $\varepsilon^{\omega^\prime,\gamma^\prime}_t$.}
\label{fig:grouping}
\end{figure}

Clearly, plugging all operators directly into formulas for $\alpha$ and $\beta$ would make the problem intractable in case of large $N$.
We show that it is possible to apply our tools effectively to the dynamics of $n \geq \max{(k,l)}$, $n \ll N $, atoms and from this analysis infer the final scaling for the whole system of $N$ atoms.
Recall that while investigating the fundamental limits to adaptive schemes we always consider the limit $t \rightarrow 0$, and the only relevant order of the dynamics we need to take into account is the linear one.
Hence, we may replace the original dynamics as represented by \eqref{eq:nonlinear} with a scheme where each $n$-tuple of particles experiences the dynamics sequentially. By a Trotter expansion argument this will introduce only a $O(t^2)$ difference due to a potential lack of commutation of the operators acting on different subsets of particles, see Fig.~\ref{fig:grouping}.
This represents the dynamics in terms o elementary operations $\varepsilon^{\omega^\prime,\gamma^\prime}_t$ acting on $n$ particles only, which we refer to as subchannels.
In the above scheme the $H$-box denotes a free unitary evolution of $k$ particles for a time $t$, while the $L$-box represent the noisy part of the dynamics lasting also for a time $t$. In order to keep the equivalence to the original problem we need to rescale the noise coefficient $\gamma^\prime = \gamma/\bi_l$, where $\bi_l = \binom{N}{n} \binom{n}{l}/\binom{N}{l} \propto N^{n-l}$. The rescaling is necessary since the number of noisy gates  applied is enhanced by a factor of $\bi_l$ compared with the original dynamics.
Similarly, we need to modify the  Hamiltonian evolution by rescaling $\omega^\prime = \omega/\bi_k\propto \omega N^{-(n-k)}$---in the example of Fig.~\eqref{fig:grouping} this is not necessary since $k=n$.

Let us assume that it is possible to find $h_\varepsilon$ that makes $\beta_\varepsilon = 0  + O(t^2)$,
 where $\beta_\varepsilon$ should be understood as $\beta$ operator corresponding to the elementary dynamics $\varepsilon^{\omega^\prime,\gamma^\prime}_t$. This again corresponds to the situation that the Hamiltonian $H$ belongs to $\mathcal{S}_\varepsilon$ which is constructed from noise operators entering $\varepsilon^{\omega^\prime,\gamma^\prime}$
 according to Eq.~\eqref{eq:space}.
 We can now apply the bound~\eqref{eq:qfiboundt} treating $\varepsilon^{\omega^\prime,\gamma^\prime}_t$ as the fundamental building block for the adaptive strategy
and  since there are now $\binom{N}{n} T/t$ such elementary blocks we arrive at:
$F_Q \leq 4 \binom{N}{n} \Vert \alpha^{(1)}_{\varepsilon^{\omega^\prime,\gamma^\prime}} \Vert T$.

Let us inspect Eqs.~(\ref{eq:beta},\ref{eq:alpha})  in order to understand the impact of the rescaling factors $\bi_l$, $\bi_k$ on the value of the above bound. Rescaling of $\gamma$ introduces an additional $1/\sqrt{\bi_l}$ factor to all $L$ operators. Taking additionally
into account that the Hamiltonian is rescaled by $1/\bi_k$ factor, then
according to $\eqref{eq:beta}$ in order to satisfy $\beta^{(1)}=0$ constraint, $\mathfrak{h}^{(0)}$ needs to be rescaled by $\bi_l/\bi_k$ while
$\mathbf{h}^{(\frac{1}{2})}$  by $\sqrt{\bi_l}/\bi_k$ factors.

Together with \eqref{eq:alpha} this implies that $\alpha^{(1)}$ is rescaled by $\bi_l/\bi_k^2$.
Therefore we finally arrive at
\begin{equation}
F_Q \leq 4 T \Vert \alpha^{(1)}_{\varepsilon^{\omega,\gamma}}\Vert\binom{N}{n} \frac{\bi_l}{\bi_k^2} \propto T N^{2k -l}.
\end{equation}

Note that the obtained scaling agrees with what we have obtained in the models discussed in Sec.~\ref{sec:losses}. In that case the situation where $\beta^{(1)}$ could be made zero corresponded to
 either $n=1, k=1, l=1$ or $n=2, k=2,l=2$, and indeed we obtained respectively $TN$, and $TN^2$ scalings.
This shows that the general approach based on splitting the complex multiparticle dynamics into small subchannels involving only the number of particles required to model a given degree of nonlinearity is sufficient to obtain a proper scaling of the precision bounds. This general approach can also be utilized to obtain quantitative bounds which will be the subject of a separate publication \cite{Czajkowski2017}.

The $N^{2k -l}$ scaling of QFI or equivalently $N^{-(k-l/2)}$ scaling of parameter estimation precision
has been also observed  in \cite{Beau2016}, for protocols based on utilizing GHZ class of states and models where all $H$ and $L_i$ operators commute. Our approach proves the scaling for both the most general class of states and adaptive strategies. Once our bound
can be derived (i.e. $\beta^{(1)}$ can be set to zero), we can claim that in models considered in  \cite{Beau2016} indeed the GHZ as well as product states provide the optimal scaling.
In the more general approach, however, where arbitrary states and adaptive strategies are allowed, this will not necessarily be the case.

To show both the power and simplicity of our approach, let us therefore consider the class of dynamics, which we will refer to as non-linear metrology with multi-particle dephasing, where  $L_i$ and $H$ commute as considered in \cite{Beau2016}, and try to apply our methods to derive general precision bounds
in this case.  Let us take
$H_\nu = \sigma_z^{\nu_1}\otimes \dots \otimes \sigma_z^{\nu_k}$ and
$L_\mu = \sigma_z^{\mu_1} \otimes \dots \otimes \sigma_z^{\mu_l}$.
Let us inspect the structure of the subspace $\mathcal{S}$, see Eq.~\eqref{eq:space},
and invoke the representation of the dynamics in terms of subchannels $\varepsilon $
as depicted in Fig.~\ref{fig:grouping} and ask for what $k$, $l$ we can satisfy
the $H \in \mathcal{S}_\varepsilon$ condition.
The obvious case is $k=l$ when we simply consider $n=k=l$ subchannels in which case the operator $H$ is proportional to the operator $L$.
Similarly we can show that $H \in \mathcal{S}$ if $k=2l$, since if we take $n=k$ we can obtain $H$ from products of $L$ acting on
two separate subsets of $l$ particles.
More generally, provided $k$ is even and $k \leq 2l$, we can consider $n= l+k/2$ particle subchannels and obtain $H$ by multiplying $L$ acting on two sets of $l$ particles where
$(l-k/2)$ of them overlap. This way the product of two $\sigma_z$ on overlapping particles produce the identity and we can obtain
a $2l - 2(l -k/2) = k$ fold tensor product of $\sigma_z$ acting on the remaining particles.

In particular for non-linear metrology $k=2$ and linear dephasing $l=1$ we automatically get $F_Q \lesssim T N^{3}$ bound while for two-body Hamiltonian $k=2$ and nonlinear dephasing $l=2$ we
get $F_Q \lesssim T N^2$, proving fundamental character of the scaling obtained in \cite{Beau2016}. However, if we take
$k=1$ and $l=2$, then according to our approach $H \notin \mathcal{S}_\varepsilon$---as we cannot obtain a single $\sigma_z$ from products of two or four $\sigma_z$ that appear in the definition of $\mathcal{S}_\varepsilon$. Note, that indeed in this case the subspace spanned by $\ket{0}^{\otimes N } \pm \ket{1}^{\otimes N}$ is actually immune to decoherence as all $L_i$ operators involve product of two $\sigma_z$ acting on different sites and therefore yield a trivial $1$ factor when acting on states from this subspace. Still the single body Hamiltonian  acts nontrivially, and we get $F_Q \propto T^2 N^2$ showing the possibility of  better than linear scaling both in $N$ and in $T$ and demonstrating that indeed in this case the bound $F_Q \lesssim T N^{2k-l} \propto T $  is invalid.

Let us also comment here on the issue of optimization of quantum metrological protocols
under the fixed definite particle number or just the mean number of particles fixed.
In linear metrology QFI scales quadratically in the decoherence-free case and hence maximization of QFI under fixed mean particle number
may lead to surprising conclusions of possibility of beating Heisenberg scaling or even reaching arbitrary high values, see e.g. \cite{Anisimov2010}. This should be understood as deficiency of the QFI figure of merit which in general provides only a lower bound on achievable uncertainty via the Cram{\'e}r-Rao bound whereas saturability of the bound requires more detailed arguments \cite{Tsang2012, Berry2012, Giovannetti2012, hall2012does}. If decoherence makes QFI scale linearly with the number of particles, though, this issue becomes non-existent
and one can replace fixed particle numbers with mean particle numbers in the formulas for the bounds \cite{Demkowicz2013, jarzyna2015true, demkowicz2015quantum}. In the non-linear models discussed above, even in presence of decoherence, the bounds will in general scale super-linearly with particle numbers and hence again the task of maximizing QFI under the fixed mean particle number will be ill-posed and as a result meaningful discussion of such problems would require going beyond the QFI paradigm.



\section{Conclusions and discussion}
Throughout this work, we have assumed Markovian semi-group dynamics. Validity of this approximation requires in particular
 coarse graining of the evolution of the system on time scales larger than the environmental characteristic relaxation time scales.
 Therefore the ,,continuous limit'' $t \rightarrow 0$, we have adopted while deriving our bounds, should be understood
 as the limit of very short times, but still in the regime where the Markovian approximation holds.
 Consequently, fundamental character of the bounds we derive, hinges upon the assumption that
 one cannot operate on time scales shorter than the ones characteristic for the Markovian approximation.
 If this assumption is dropped one may expect different time scalings of QFI, see e.g. \cite{Chin2012, Smirne2016, Sekatski2016}.
Still, combining the framework of the most general adaptive protocols with non-Markovian or even Markovian non-semigroup dynamics, is a non-trivial task. In particular, focusing solely on the reduced system dynamics may not be in general sufficient to describe the effects of control operations acting on the system, and more information on system+environment interaction may be required. Moreover, in such scenarios, one cannot a priori justify the framework where the time used up by control operations is not relevant, as well as argue that a model with infinitely many control operations is the optimal one. We feel that providing a general framework to study the potential of adaptive quantum metrological protocols in presence of non-Markovian noise goes beyond direct generalization of the methods of this paper and leave this task for future research.

Another way of generalizing results of this work is to assume the environment can be partially monitored, which may open up completely new possibilities of fighting decoherence. It is also interesting to study multiparameter estimation scenarions were adaptiveness seems to play some role already at the decoherence-free level \cite{yuan2016ultimate}.

To conclude. We have provided a simple algebraic criterion, Eq.~\eqref{eq:space}, determining at most linear time scaling of QFI in estimation of a unitary parameter under general Markovian dynamics. Its high utility stems from the fact that it deals directly  with the Hamiltonian and noise operators appearing in the Master equation and does not require integration of the dynamics. We have shown how the bounds can be derived in atomic interferometric models involving many-body interactions.
In the qubit case it has been shown that the condition \eqref{eq:space} is sufficient and necessary for fundamental linear time scaling of QFI. In case of arbitrary dynamics in arbitrary dimension, however, it has only been proven to be a sufficient condition. Whether it is also a necessary one remains an open question.

\emph{Note added.} After completion of this work, a paper \cite{zhou2018achieving} appeared, where apart from independent derivation of the results presented in Sec.~\ref{sec:linear} of our paper, an explicit construction of quantum error-correction protocol has been provided yielding $T^2$ scaling of the QFI whenever condition \eqref{eq:space} is not satisfied.
As a result, this paper answered the open problem stated above on whether the condition
\eqref{eq:space} is indeed the \emph{if and only if} condition for impossibility of preserving the $T^2$ scaling of QFI via the most general quantum  protocols in the presence of noise for systems of arbitrary dimension. Note also that our conditions for an effective error correction protocol  (\ref{eq:ecconditions}-\ref{eq:conditions1}) are equivalent to the ones provided in \cite{zhou2018achieving}. In particular by expressing projector $\Pi_C$ from \cite{zhou2018achieving} using $\ket{\phi}$, $\ket{\xi}$ from Sec.\ref{sec:quadratic} of our paper as $\Pi_C = \ket{\phi}\bra{\phi} + \ket{\xi}\bra{\xi}$, we see that their condition (15) is equivalent to ours (\ref{eq:conditions2}-\ref{eq:conditions1})
while their condition (17) is equivalent to ours \eqref{eq:ecconditions}---on the one hand one can always find two states in this subspace $\Pi_C$ that are coupled by the Hamiltonian, on the other if two states from the subspace are coupled the action of the Hamiltonian is nontrivial.
Finally, let us briefly describe how the universal error correction code of \cite{zhou2018achieving} appears in our language. The key idea is to use the spectral decomposition of the part of the Hamiltonian perpendicular to the noise space $H_\perp = P-Q$, where $P$ and $Q$ are orthogonal, positive semi-definite and $\tr \,P = \tr \,Q$ since $H_\perp$ is traceless by construction. Next, on the probe plus ancilla system define the corresponding purifications states $\ket{p}$ and $\ket{q}$ satisfying
\be
P = h_\perp \tr_A \proj{p} \quad \text{and} \quad Q = h_\perp \tr_A \proj{q},
\ee
where $h_\perp >0$ since $H_\perp$ is non-zero. Then the virtual qubit states of Sec.~\ref{sec:quadratic} can be defined as
$\ket{\phi} = \frac{1}{\sqrt{2}}\left(\ket{p}+\ket{q}\right)$ and $\ket{\xi} = \frac{1}{\sqrt{2}}\left(\ket{p}-\ket{q}\right)$.

\section*{Acknowledgements}
We thank Wolfgang D\"ur, Janek Ko{\l}ody{\'n}ski, Animesh Datta, Philipp Treutlein and Krzysztof Paw{\l}owski for fruitful discussions.
This work was supported by the Polish  Ministry  of  Science  and  Higher
Education  Iuventus  Plus  program  for  years  2015-2017
No. 0088/IP3/2015/73, National Science Center (Poland) grant No. 2016/22/E/ST2/00559 and Swiss National Science Foundation grant P300P2\_167749.

\appendix

\section{Calculating the bound via semi-definite programming}
\label{sec:sdp}
Given the condition $H \in \mathcal{S}$ is satisfied, we know that the scaling of QFI is bound to
be linear in $T$. In order to obtain the tightest bound of the form \eqref{eq:alphabound}
we need to minimize the operator norm of $\alpha^{(1)}$ keeping the constraint $\beta^{(1)}=0$.
First we construct a matrix $A$
\begin{equation}
A = \mat{cc}{
\sqrt{\lambda} \openone & \mathbf{h}^{(\frac{1}{2})\dagger} \openone + \mathbf{L}^\dagger \mathfrak{h}^{(0)}  \\
\mathbf{h}^{(\frac{1}{2})} \openone + \mathfrak{h}^{(0)} \mathbf{L} & \sqrt{\lambda} \openone^{\otimes J}   }.
\end{equation}
Minimizing the operator norm $\Vert \alpha^{(1)} \Vert$ is now equivalent to minimizing $\lambda$ subject to
$A \geq 0$ with the additional constraint coming from the equation $\beta^{(1)}=0$.
The QFI bound can therefore be written as
\begin{align}
\begin{split}
F_Q \leq  4 T & \min_{\{h_{00}^{(1)},\mathbf{h}^{(\frac{1}{2})},\mathfrak{h}^{(0)}\}} \lambda, \\
& \textrm{subject to: } A\geq 0,\  \beta^{(1)}=0.
\end{split}
\end{align}
The problem of determining the bounds is now fully specified as a semi-definite program using only the operators appearing in the Master equation.

\section{Tightening the bound using physical constraints}
\label{sec:constraints}
The bound in the form \eqref{eq:qfiboundt}, which involves operator norms, may in some cases be tightened by taking into account additional physical
constraints present in the considered problem.

\subsection{Superselection rules}

It  is often the case in practice that the experimentalist does not have access to operations that create coherence between some sectors of a the total Hilbert space. For example, in no photonic experiment one is able to create, manipulate or detect coherence between sectors of different total number of particles $n_\t{t}=a_1^\dag a_1 + a_2^\dag a_2 +...$ in all modes (including local oscillators). This is a consequence of symmetries, like the time translation symmetry enforcing energy conservation in the example with particles. It is usually the case that the evolution, Eq.~\eqref{eq:ME}, does not create such coherences either, unless the evolution is ``active'' and describes the interaction of the system with a source. This can be formalized by introducing a map
\be
\mathcal{P}(\cdot) =\sum_{k} P_k \cdot P_k
\ee
with $P_k P_j =\delta_{k,j}P_k$, that projects the state of the systems onto sectors that are eigenspaces of the conserved quantities labeled by $k$.  Control operations and evolution are incoherent  they commute with map $\mathcal{P}$, implying for instance that $H$ is block diagonal $\mathcal{P}(H)=H$.

Under such circumstances QFI of a state $\rho$ might be overestimating the extractable classical Fisher information of the estimated parameter, as the information might be, for example, encoded in the coherence between different sectors. To get a tight expression one then has to explicitly account for the conservation law, by looking on QFI of the incoherent state $\mathcal{P}(\rho)$. A similar treatment can be done in our case. Since the projection map $\mathcal{P}\circ\mathcal{P}=\mathcal{P}$ commutes with both the controls and the evolution and has to be applied on the final state, we can explicitly account for the conservation law by considering the process $\tilde \cE_\omega(t)$ where the free evolution induced by the master equation~\eqref{eq:ME} is interjected with $\mathcal{P}$  after each infinitesimal time step $t$
\be
\rho_\omega(t_0+t) =\tilde\cE_\omega(t)\big(\rho(t_0) \big)= \mathcal{P}\circ \cE_\omega(t)\circ \mathcal{P}\big(\rho(t_0)\big).
\ee
This composed channel $\tilde\cE_\omega(dt)$ has a different set of Kraus operators
\begin{align}
K_{0;k} &= P_k - P_k\left(\frac{1}{2} \mathbf{L}^\dagger \mathbf{L} + i H \omega \right)P_k t + O(t^{2}), \\
K_{j;k\ell} &= P_k L_j P_\ell \sqrt{t} + O(t^{3/2}).
\end{align}
But similarly as before we can get the condition for the linearity of QFI as
\begin{align}
\beta^{(1)} = H&+  \sum_{k} h_{00;k}^{(1)}P_k + {\mathbf{h}_{k}}^{\dagger (\frac{1}{2})} P_k\mathbf{L}P_k +P_k \mathbf{L}^\dagger P_k {\mathbf{h}_{k}}^{({\frac{1}{2}})}\nonumber \\
&+ \sum_{k,\ell} P_k \mathbf{L}^\dagger \mathfrak{h}_{k\ell}^{(0)} P_\ell \mathbf{L} P_k.
\end{align}
Note that because $H$ is block-diagonal, we set all the entries of $h$ leading to terms of the form $P_k \cdot P_\ell$ with $(k\neq \ell)$ to zero. Hence, when the process in incoherent as described by the map $\mathcal{P}$ QFI scales linearly if each $H_k =P_k H P_k$ satisfies
\begin{align}
\begin{split}
H_k \in \t{span}\{P_k, (P_k L_j P_k)^H,\ii (P_k L_j P_k)^{AH},\\
 (P_k L_j^\dag P_\ell L_i P_k)^H, (P_k L_j^\dag P_\ell L_i P_k)^{AH}\}
 \end{split}
\end{align}
for all $\ell,j$ and $i$.

Analogously, the minimization of the operator norm of $\alpha^{(1)}$ should be done for
a block diagonal operator:
\begin{align}
\begin{split}
\alpha^{(1)} = \sum_{k\ell}& \left(\mathbf{h}_k^{(\frac{1}{2})} P_k + \mathfrak{h}_{k\ell}^{(0)}P_{\ell}\mathbf{L}P_k \right)^\dagger \\
&\times \left( \mathbf{h}_k^{(\frac{1}{2})} P_k + \mathfrak{h}_{k\ell}^{(0)}P_{\ell}\mathbf{L}P_k\right ).
\end{split}
\end{align}

\subsection{Restriction to a subspace}
The most obvious situation, is when the system state is restricted to live in a particular subspace of the Hilbert space, and neither the Hamiltonian nor the noise operators $L_i$, nor the active
control operations move the state out of it.
In this case all operator norms appearing in \eqref{eq:qfiboundt} as well as conditions
on vanishing operators $\beta$ should be understood as restricted to this subspace.
This is for example the case, when we deal with atomic or optical systems with fixed number of particles.
Formally, let $P$ be the projection on the subspace.
In all the formulas involving $\beta$ and $\alpha$ operators, we should simply replace
all $H$ and $L_i$ by respectively $P H P$ and $P L_i P$. This situation can be viewed as a special case of the super-selection rule situation when we leave only a projection on a single subspace.

\subsection{Fixing the number of particles in the protocol involving losses}
Let us consider the situation, where we start with a state with a fixed number of particles
which experiences losses as in the case of models discussed in Sec.~\ref{sec:losses}.
If we start with an $N$ atom state, then clearly due to losses we will end up with a mixture of states with different atom numbers.
Recall, however, that we work in the most general adaptive metrological scenario. We will therefore assume that
at each adaptive step we feed the system back with the lost atoms and carry on the evolution with an $N$-atom state.
In practice this would amount to performing a non-demolition measurement of the number of remaining atoms, and
then since every state of $n<N$ atoms can be isomorphically transcribed to the state of $N$ atoms we do not lose any
parameter information that was potentially present in the state as a result of earlier dynamics.
With this in mind, and recalling that adaptive steps in the most general strategy can be chosen to be infinitesimally small, we can think of this situation as effectively an evolution with fixed number of particles.

\subsection{Taking into account the state-time dependence}
Let us consider the situation, where we have some additional knowledge of the form of the state
of the system as it evolves in time under  our protocol.
Let us denote the general error-correction assisted evolution by a CPTP map $\mathcal{C}^\omega_t$.
 Let the initial state of the system plus ancilla be $\varrho_0=\proj{\psi_0}$ \, then the state at time $T$ is given by
\be
\varrho_T^\omega = \mathcal{C}^{\omega}_T \left(\varrho_0\right)=\sum_{\bf i} K_{\bf i} \proj{\psi_0} K_{\bf i}^\dag,
\ee
with some Kraus operators $K_{\bf i}$. We assume here the error-correction protocol does not depend on time, but this is not crucial for the results that follows. QFI of any such channel satisfies
\be\label{eq:QFIgen}
F_Q \leq 4\min_{\tilde {\bm K}}\tr \sum_{\bf i} \dot {\tilde K}_{\bf i} \proj{\psi_0} \dot {\tilde K}_{\bf i}^\dag,
\ee
with the minimization over all Kraus representations of the channel $\tilde {\bf K} \simeq {\bf K}$.
As we argued before the channel can be decomposed into a product of infinitesimal channels $\mathcal{C}^{\omega}_T=\bigcirc_{n=1}^{m=T/dt}\mathcal{C}^{\omega}_{dt}$, so that a Kraus operator of the global overall process $K_{\bf i} = \Pi_{n=1}^{m} K_{i_n}$. So the derivative $\dot K^{T/dt}$ gives rise to a sum of $m$ terms, with each term having a derivative on an individual $\dot K_{i_n}$. This observation together with the trace preserving propeperty of $\mathcal{C}^{\omega}_{t}$ allows us to rewrite the rhs of Eq.~$\eqref{eq:QFIgen}$ as
\begin{align}
&4\int dt' \tr \sum_{i_{t'}}\dot {\tilde K}_{i_{t'} }\cE_{\omega, t'}(\varrho_0) \dot {\tilde K}_{i_{t'} }^\dag \nonumber\\
&+ 4\int_{t'>t'' }\!\!\!dt' dt''\tr \sum_{i_{t'}}\dot {\tilde K}_{i_{t'} }\mathcal{C}^{\omega}_{t'-t''}\Big( \sum_{i_{t''}} {\tilde K}_{i_{t''} } \mathcal{C}^{\omega}_{t''}(\varrho_0) \dot {\tilde K}_{i_{t''} } \Big) {\tilde K}_{i_{t'} }^\dag\nonumber\\
&+h.c.
\end{align}
Choosing the local Kraus representation ${\bf K}$ such the the corresponding $\beta^{(1)}$ is set to zero at the order $dt$ (if this is possible) sets the two last terms to zero, and gives the bound
\be
F_Q \leq 4\int_{0}^T dt \tr\left(\sum_{i_{t}} \dot {\tilde K}_{i_{t} }^\dag \dot {\tilde K}_{i_{t} }\right) \mathcal{C}^{\omega}_{t}(\varrho_0).
\ee
The term in the paranthesis equals to  $\alpha^{(1)}$,  so that we just obtained a state dependent upper bound on QFI.
\be\label{eq:stb}
F_Q \leq 4 \int_{0}^T dt \min_{\tilde {\bf K}\, \t{s.t.}\, \beta^{(1)}=0}\tr \, \alpha^{(1)}(t) \, \rho_{t}.
\ee
Note that this is in accordance with the general (state independent) bound, as for any state $\rho$ it is true that $\tr \,\alpha^{(1)} \rho \leq \lVert  \alpha^{(1)}\rVert$. Moreover, if one knows some properties of the time evolution of the state $\rho_{t}$ under the error-correction assisted protocol, the bound in Eq.~\eqref{eq:stb} can be much tighter than the generic one $\lVert\alpha^{(1)}\rVert T$.

To give a specific example of utility of such a bound, consider atomic interferometry in presence of single-particle
 losses as discussed in Sec.~\ref{sec:losses}. While deriving the bound, we assumed that lost atoms can be replaced with new ones.
 One might consider, however, a more realistic situation when this is not the case and the lost atoms are not being replaced.
For simplicity, let us assume that losses are state independent $\gamma_1 = \gamma_2 = \gamma$. This means that starting with a state of $N$ atoms, after time $t$ we will have an average number of atoms equal to $N e^{- \gamma t}$.
Applying the time dependent bound, and using \eqref{eq:bound1losses} (which is also valid for incoherent mixtures of states
with different number of atoms, where the mean number is $N$), we can therefore write:
\be
F_Q \leq  \int_{0}^t dt' \frac{N e^{-\gamma t}}{\gamma} = \frac{N}{\gamma^2}(1 - e^{-\gamma T}),
\ee
which is in general tighter than $F_Q \leq  NT / \gamma$.

\section{Error-correction scheme}

\subsection{Canonical form of noise operators}
\label{sec:noise operators}

If any of the noise operators $L_j$ is not traceless, it can be decomposed into $L_j = \lambda \openone + \bar L_j$, with $\tr \openone \bar L_j =0$. Under this substitution the Lindblad operator remain the same (now denoted by $\bar L_j$) but an additional Hamiltonian term $-\ii \left[ \frac{\ii}{2}(\lambda^* \bar L_j-\lambda \bar L_j^\dag) ,\rho\right]$ appears in the master equation, that can be compensated by control operations. In addition, for any set of noise operators $\{L_j\}$ it is always possible to find an equivalent, i.e. giving rise to the same dynamics, set $\{\bar L_j\}$ of operators that are orthogonal \cite{Pearle2012} under the Hilbert-Schmidt product. Hence, one can always put the noise operators in the form $\tr L_j  =0$ and $\tr L_k^\dag L_j=0$ for all $j$ and $k$.

\subsection{Derivation of error-correction conditions (\ref{eq:ecconditions}-\ref{eq:conditions1})}
\label{sec:errorcorrection}

 A general map $C$ that maps the system from the big Hilbert space $\cH$ back onto a qubit subspace $\cH_Q$, can be written as
\be
C(\cdot) = \sum_\ell R_\ell \cdot R_\ell^\dag,
\ee
with $R_\ell = \mu_\ell \coh{\phi}{\Phi_\ell} + \lambda_\ell \coh{\xi}{\Xi_\ell}$ satisfying $\openone_\cH =\sum_\ell R^\dag_\ell R_\ell = \sum_\ell \big(|\mu_\ell|^2\, \proj{\Phi_\ell} +|\lambda_\ell|^2\, \proj{\Xi_\ell}\big)$. For our purpose it is natural to consider the case where the two subspaces $\cH_\Phi =\t{span}\{\ket{\Phi_\ell}\}$ and $\cH_\Xi=\t{span}\{\ket{\Xi_\ell}\}$ are orthogonal, and the operator $\Pi_\Xi= \sum_\ell |\lambda_\ell|^2\, \proj{\Xi_\ell}$ and $\Pi_\Phi= \openone_\cH -\Pi_\Xi$ are orthogonal projectors. The map is then fully specified by the projector $\Pi_\Xi$  and the operator $M = \sum_{\ell} \mu_\ell \lambda_\ell^* \coh{\Xi_\ell}{\Phi_\ell}$. Given that $\t{dim}(\cH_\Xi)\leq \t{dim}(\cH_\Phi)$, it is possible to find a Kraus representation of the map $C$ for which the vectors $\lambda_\ell^* \ket{\Xi_\ell}$ are orthonormal, hence we assume this in the following (the same holds for $\mu_\ell^* \ket{\Phi_\ell}$ in the case $\t{dim}(\cH_\Xi)> \t{dim}(\cH_\Phi)$). Recall that we require from the error-correction scheme to preserve the qubit subspace spanned by $\ket{\phi}$ and $\ket{\xi}$ under the action of the noise $\mathcal{L}$. For the map $C$ this implies
\begin{align}
\tr \Pi_\Xi \big(\proj{\phi} + dt \mathcal{L}(\proj{\phi})\big) &= 0\\
\tr \Pi_\Xi \big(\proj{\xi} + dt \mathcal{L}(\proj{\xi})\big) &=1 \\
\tr \Pi_\Xi \big(\coh{\phi}{\xi} + dt \mathcal{L}(\coh{\phi}{\xi})\big) &=0 \\
\tr M \big(\proj{\phi} + dt \mathcal{L}(\proj{\phi})\big) &= 0\label{eq:C4}\\
\tr M \big(\proj{\xi} + dt \mathcal{L}(\proj{\xi})\big) &= 0\label{eq:C5}\\
\tr M \big(\coh{\phi}{\xi} + dt \mathcal{L}(\coh{\phi}{\xi})\big) \label{eq:C6}&= 1
\end{align}
The first three equations imply that $\ket{\phi}\in\cH_\Phi$ and  $\ket{\xi}\in\cH_\Xi$, as well as $L_k\ket{\phi}\in \cH_\Phi$ and $L_k\ket{\xi}\in \cH_\Xi$ for all $k$. Eqs.~\eqref{eq:C4} and \eqref{eq:C5} impose in addition $\bra{\phi}{\bf L}^\dag {\bf L} M \ket{\phi}=$ $\bra{\xi}M{\bf L}^\dag {\bf L} \ket{\xi}=0$. Finally, Eq.~\eqref{eq:C6} requires that $\tr M \coh{\phi}{\xi}=1$, allowing to write $M =\coh{\xi}{\phi}+M_\perp$, and
\begin{align}
\label{eq:cond}
\begin{split}
\tr M & \sum_k L_k \coh{\phi}{\xi} L_k^\dag \\
& - \frac{1}{2}\left( \bra{\xi}{\bf L}^\dag {\bf L}\ket{\xi}  +\bra{\phi}{\bf L}^\dag {\bf L}\ket{\phi}\right)=0.
\end{split}
\end{align}
This condition can be satisfied if and only if there is a unitary $U$ relating all the vectors $L_k \ket{\phi}$
to $L_k \ket{\xi}$. In which case setting $M=U$ satisfies Eq.~\eqref{eq:cond}. In turn such a unitary exists if and only if one has
\be
\bra{\phi} L_k^\dag L_j \ket{\phi}=\bra{\xi} L_k^\dag L_j \ket{\xi}\quad \forall\, j,k,
\ee
i.e. the Gramm matrices for the two vector sets are the same.

In summary, an error correction scheme that satisfies all of the requirement above exists if and only if one can find two states $\ket{\phi}$ and $\ket{\xi}$ such that
\begin{align}
&\t{(a)}\quad\bra{\phi} H\ket{\xi}\neq 0\\
&\t{(b)}\quad\bra{\phi} L_k^\dag L_j \ket{\xi}=\bra{\phi} L_j \ket{\xi}=0\\
&\t{(c)}\quad\bra{\phi}L_k^\dag L_j \ket{\phi}= \bra{\xi}L_k^\dag L_j \ket{\xi}.
\end{align}
for all $k$ and $j$. Where the condition (b) and (c) are properties of error correction, while (a) has to be satisfied in order to keep non-trivial  unitary evolution in the qubit subspace $\cH_Q$.

\section{Derivation of the fundamental precision bound in non-linear metrology with two-body losses}
\label{sec:twobody}
Starting from \eqref{eq:bound2lossesder} and substituting $a_1^{\dagger 2}a_1^2$ with $n(n-1)$ and $a_2^{\dagger 2}a_2^2$ with $(N-n)(N-n-1)$,
calculation of the operator norm amounts to maximization over $0 \leq n \leq N$. The bound \eqref{eq:bound2lossesder}  can be therefore rewritten as
\be
F_Q \leq \min_{\xi} \max_{n }\frac{T}{4} \left[(\xi -1)^2 A + (\xi+1)^2 B \right ],
\ee
where
\begin{align}
\begin{split}
&A = \frac{n(n-1)}{\gamma_{11}} + \frac{(N-n)(N-n-1)}{\gamma_{22}}, \\
&B= \frac{4(N-n)n}{\gamma_{12}}.
\end{split}
\end{align}
Performing minimization over $\xi$, yields $\xi = (A-B)/(A+B)$ and consequently
\begin{align}
\begin{split}
& F_Q  \leq  \frac{N^2 T}{\gamma_{12}} \max_{0 \leq x \leq 1} \bigg[\frac{1}{4 (1-x)x} \\
 &+ \frac{1}{\frac{\gamma_{12}}{\gamma_{11}}(x-\frac{1}{N})x+ +\frac{\gamma_{12}}{\gamma_{22}}(1-x)(1-x-\frac{1}{N})}   \bigg]^{-1}
\end{split}
\end{align}
where we introduced $x=n/N$, $0 \leq x \leq 1$. In case $\gamma_{11}=\gamma_{22}$ this maximization can be performed analytically and results in
\begin{equation}
F_Q \leq \frac{N^2 T}{\gamma_{12}} \begin{cases}
\frac{2(1-1/N)}{(1+\sqrt{\lambda})^2},  & \lambda  \geq 1-\frac{2}{N}  \\
\frac{1}{1+\lambda \frac{N}{N-2}}, & \lambda < 1-\frac{2}{N}
\end{cases},
\end{equation}
where $\lambda = 2 \gamma_{11}/\gamma_{12}$ and yields \eqref{eq:bound2losses} in the $N \gg 1$ limit.
In a general case $\gamma_{11} \neq \gamma_{22}$ the maximization over $x$ can be easily performed numerically.

%

\end{document}